\documentclass[runningheads]{llncs}
\usepackage{graphicx}
\usepackage{xcolor}
\usepackage{csquotes}
\usepackage{comment}
\usepackage[acronym]{glossaries}
\usepackage{blindtext}
\usepackage{cite}
\usepackage{amsmath}
\usepackage{amsfonts}
\usepackage{hyperref}
\usepackage{lmodern}

\newacronym{ai}{AI}{artificial intelligence}
\newacronym{ias}{IAS}{Intelligent Assistive Systems}
\newacronym{pwd}{PwD}{Persons with Dementia}
\newacronym{iat}{IAT}{intelligent assistive technology}
\newacronym{ecq}{ECQ}{ethical compliance quantification}
\newacronym{simdem}{SimDem}{SimDem – A Multi-agent Simulation Environment to Model Persons with Dementia and their Assistance}

\newcommand{\nhelp}{$\mathit{n_{help}}$}
\newcommand{\te}{$\mathit{TE}$}

\makeatletter
\renewcommand\section{\@startsection{section}{1}{\z@}%
                       {-8\p@ \@plus -4\p@ \@minus -4\p@}%
                       {6\p@ \@plus 4\p@ \@minus 4\p@}%
                       {\normalfont\large\bfseries\boldmath
                        \rightskip=\z@ \@plus 8em\pretolerance=10000 }}
\renewcommand\subsection{\@startsection{subsection}{2}{\z@}%
                       {-8\p@ \@plus -4\p@ \@minus -4\p@}%
                       {6\p@ \@plus 4\p@ \@minus 4\p@}%
                       {\normalfont\normalsize\bfseries\boldmath
                        \rightskip=\z@ \@plus 8em\pretolerance=10000 }}
\renewcommand\subsubsection{\@startsection{subsubsection}{3}{\z@}%
                       {-4\p@ \@plus -4\p@ \@minus -4\p@}%
                       {-1.5em \@plus -0.22em \@minus -0.1em}%
                       {\normalfont\normalsize\bfseries\boldmath}}
\makeatother
\begin{document}
\title{Towards Measuring \enquote{Ethicality} of an Intelligent Assistive System} 
\author{
M. Salman Shaukat\inst{1} \and
J.-C. Põder \inst{2} \and
Sebastian Bader\inst{1} \and
Thomas Kirste \inst{1}}
\authorrunning{Shaukat et al.}
\institute{Computer Science Department, University of Rostock, Germany \and Faculty of Theology, University of Rostock, Germany}
% Faculty of Theology, University of Rostock
\maketitle              % typeset the header of the contribution
\begin{abstract}
Artificial intelligence (AI) based assistive systems, so called \acrfull{iat} are becoming increasingly ubiquitous by each day. \acrshort{iat} helps people in improving their quality of life by providing intelligent \textit{assistance} based on the provided data. Few examples of such \acrshort{iat}s include self-driving cars, robot assistants and smart-health management solutions. However, the presence of such autonomous entities poses \textit{ethical challenges} concerning the stakeholders involved in using these systems. There is a lack of research when it comes to analysing how such \acrshort{iat} \textit{adheres} to provided ethical regulations due to ethical, logistic and cost issues associated with such an analysis. In the light of above-mentioned problem statement and issues, we present a method to measure \enquote{ethicality} of an assistive system. To perform this task, we utilised our simulation tool that focuses on modelling navigation and assistance of \acrfull{pwd} in indoor environments. By utilising this tool, we analyse how well different assistive strategies adhere to provided ethical regulations such as \textit{autonomy}, \textit{justice} and \textit{beneficence} of the stakeholders. 

\keywords{Artificial Intelligence \and Ethical AI \and Healthcare}
\end{abstract}

%%%%%%%%%%%%%%%%%%%%%%%%%%%%%%%%%%%%%%%%%%%%%%%%%%%%%%%%%%%%%%%%%%%%%%%%%%%%%%%%%%%%%%%%%%%%%%%%%%%%%%%%
\section{Introduction}
\subsection{IAT \& Ethics' Compliance}
Making sure that the operation of intelligent assistive technology (\acrshort{iat}) complies with ethical values is important for technology acceptance. Objective of this paper is (i) to propose the use of behavior simulation as a means for quantifying an \acrshort{iat}'s compliance with a set of ethical values, (ii) to show how this method may be used to obtain quantitative statements on an \acrshort{iat} strategy's compliance with different ethical values (\textit{ethical compliance quantification}, ECQ), and (iii) to discuss some of the issues in setting up such an evaluation model.

Intelligent assistance based on mobile and wearable devices or smart environment technologies spreads throughout our everyday life. Such \acrshort{iat}s employ methods from \acrfull{ai} that enable the assistive systems to act autonomously for their users. It is well established that the use of AI in IAT gives rise to a wide range of ethical concerns (see for instance \cite{russell2019human} for an overview).

The use of \acrshort{iat} is also being considered in the area of dementia care (see for instance the reviews \cite{ienca2017intelligent} and \cite{teipel2018use}). The user group here is highly vulnerable, creating additional ethical concerns. It is here of specific interest, that a recent study has revealed that professional stakeholders in dementia care may disagree on the interpretation of ethical values and the appropriate strategy for resolving value conflicts \cite{wangmo2019ethical}. In this paper, we focus on a specific application example in this domain, namely the use of \acrshort{iat} in nursing homes to help resident \acrshort{pwd} in orientation and wayfinding. 

When discussing the ethical challenges of employing \acrshort{iat}, there are three aspects to consider:

\begin{itemize}
    \item The relevant set of ethical values and their preference structure (which may differ across stakeholders).
    \item The mechanisms used to embed the consideration of ethical values into the decision deliberation of the \acrshort{iat}.
    \item The actual ethical compliance quantification of the \acrshort{iat}'s concrete actions in use.
\end{itemize}
The last point is specifically interesting, as it looks at evaluating what the \acrshort{iat} \textit{does} rather than at considering how to model what the IAT \textit{should do}. Note that the last point indeed has a scope beyond \acrshort{iat}, as, by focusing on the systems \textit{actions} rather than on its \textit{intentions}, it is not affected by the \acrshort{iat}'s decision making strategy and thus also applicable to assistive strategies that can be created without the use of AI technology (such as helping nursing home residents in wayfinding through appropriate architectural design).
\subsection{Simulation Based Compliance Evaluation}
The challenge is that \textit{observing} and \textit{evaluating} an assistive technology's adherence to a value set is expensive (due to the experimental logistics required) and ethically questionable (as these are experiments with human subjects). We therefore propose to use \textit{behavior simulation} for the \acrshort{ecq} of an \acrshort{iat} with respect to a set of ethical values. In this paper we present a first case study on how such a simulation-based approach to the \acrshort{ecq} of an \acrshort{iat} can be set up.

Such an approach is also interesting as it separates the \textit{decision making strategy} and the \textit{ethical value system} used in \acrshort{iat} design and in \acrshort{iat} runtime decision making from the \acrshort{ecq}. Research on computational models for ethical decision making – required for enabling machines to make such decisions – is ongoing. Various methods in diverse domains are already proposed to integrate ethical values into AI-based decision mechanisms and can be roughly categorised as: (i) Probabilistic \cite{abel2016reinforcement}, (ii) Learning‐based \cite{schramowski2020moral}, (iii) Logic‐based \cite{benzmuller2020designing}, and (iv) crowd-sourcing \cite{anderson2005medethex} approaches. At the same time, there is no consensus on a system of ethical values and its preference structure (and, by Arrow's impossibility theorem \cite{arrow1950difficulty}, such a structure may not exist). Rather than focusing on finding the \textit{Right Way} for incorporating ethical values into mechanical decision making, a simulation-based \acrshort{ecq} allows to quantify \acrshort{iat} value compliance across a \textit{range} of preference structures, thus enabling a \textit{sensitivity} analysis of the \acrshort{iat}'s strategy with respect to different individual preference structures. 

The further structure of this paper is as follows:
In section 2 we outline the rationale behind ethical value system evaluation and in section 3 we briefly introduce our behaviour simulation model \textit{SimDem} \cite{Shaukat2021SimDem}  and justify its usage for this particular study. In section 4 we explain ethical value quantification for different ethical values and present the results. Finally in section 5 we outline the main benefits and limitations of our approach and proposed future work to deal with said limitations.

%%%%%%%%%%%%%%%%%%%%%%%%%%%%%%%%%%%%%%%%%%%%%%%%%%%%%%%%%%%%%%%%%%%%%%%%%%%%%%%%%%%%%%%%%%%%%%%%%%%%%%%%
%%%%%%%%%%%%%%%%%%%%%%%%%%%%%%%%%%%%%%%%%%%%%%%%%%%%%%%%%%%%%%%%%%%%%%%%%%%%%%%%%%%%%%%%%%%%%%%%%%%%%%%

\section{Evaluating Ethical Value Systems}
\begin{comment}
\begin{itemize}
\item Ethical Perspective (Justification of our approach from an ethical perspective)
    \item Possibly outlining the wide range of different value systems, explaining the
infeasibility of arriving at a unified system, justifying an empirical approach to
ethics evaluation
\item  Also justifying the need for stakeholder-value-driven evaluation of IAS
 (independent of the IAS decision machinery)
 \item  Finally suggesting the Bioethics values as one value set of interest, to start a
 feasibility study in ECQ with
\end{itemize}
\end{comment}
The \acrshort{ecq} can be seen as an innovative addition to the methodological toolset of different well-established strategies to ethically assess and shape technological developments, such as Technology Assessment (TA) or Value Sensitive Design (VSD). ECQ can be used in the design process to test the actual compliance of a system with an intended value-set. However, and importantly, it can also be used to measure the ethicality of already existing assistive systems by testing their adherence to provided ethical regulations.

Ethical value systems used in \acrshort{iat} design can significantly vary, incorporating different values and preference structures. This situation mirrors the fact that while there is a number of historically influential ethical theories and value systems, there is no universal consensus or agreement on which approach is the only correct or the best. In machine ethics and \acrshort{iat} design, one can find a wide range of attempts to integrate different value systems into AI-based decision mechanisms, such as top-down (deontology, utilitarianism), bottom-up (casuistry) and mixed approaches to ethical values and decision making \cite{gordon2020building}. 

This situation poses challenges concerning the ethically responsible use and implementation of \acrshort{iat}. There is a need for critical evaluation of value systems used in \acrshort{iat} design and of how such \acrshort{iat} adhere to ethical regulations in a specific domain. Due to their institution-like character, IAS have potential deeply to affect and regulate our social interactions and values \cite{orwat2010software}. Making sure that \acrshort{iat} meets the best ethical standards is especially important in the field of health care and nursing which involves highly vulnerable user groups. Ethical questions should here not only be regarded but should constitute the focal point for implementing of IAS \cite{grunwald2020gestaltung}. 

On healthcare domain, the ECQ can be seen as a novel method of Health Technology Assessment (HTA) to ensure the ethicality of a technology. However, ethical pluralism observable in implementing ethics into intelligent systems poses a challenge also to technology assessment. HTA can avoid the normative task of identifying an ethical framework or value-set for ethics evaluation by relying on stakeholder-based evaluation. From an ethical perspective, the dialogical and participatory involvement of stakeholders, their values and needs, can be seen as highly important to ethics evaluation. At the same time, solely stakeholder-based evaluation seems ethically insufficient or unsatisfactory. This is so because the \textit{factual} value systems of the stakeholders do not need to be \textit{normatively} valid or acceptable. In addition, important human and moral values can also have relevance independently of whether particular stakeholders uphold them \cite{brey2010values}.

To respond to this issue, (health) technology assessment can, for example, combine different ethical theories \cite{harris2011ethical} or adapt the approach of the so-called middle principles which operates with a set of ethical values or principles which are located between high-level ethical theories and low-level particular judgements \cite{droste2003methoden}. In the field of health care, the most prominent such mid-level theory is the principlism of Tom Beauchamp and James Childress \cite{beauchamp2019principles}. They operate with four ‘principles of biomedical ethics’: \textit{autonomy}, \textit{non-maleficence}, \textit{beneficence} and \textit{justice}, which need to be carefully specified and \textit{balanced} in specific situations. Beauchamp and Childress hold that these principles or values are universal and belong to the core of our universal ‘common morality’. This value set will serve in the following as a proof concept for the ECQ presented in this paper, although our approach can be adapted also to stakeholder defined rules.

The ECQ is an innovative methodological contribution to technology assessment as it develops an empirical and quantitative approach to ethics evaluation. For decades, ethics evaluation in the field of medicine and health care technology relied more on philosophical methods such as conceptual analysis. Despite of the increasing popularity and need of empirical and interdisciplinary research within applied ethics \cite{baumann2011empirische}, there are specific challenges in quantifying approaches to qualitative phenomena such as values or quality of life. Quantifying approach relies to a significant extent on isolating, standardizing and decontextualizing strategies in defining values and value measurements. Such an operationalization has methodological advantages (objectivity, reliability, etc.), however, it  also requires (self-)critical awareness in regard to the complexity of values and ethically relevant situations. Our strategy and possible challenges in ethical value quantification will be addressed in sections 4 and 5. 
%%%%%%%%%%%%%%%%%%%%%%%%%%%%%%%%%%%%%%%%%%%%%%%%%%%%%%%%%%%%%%%%%%%%%%%%%%%%%%%%%%%%%%%%%%%%%%%%%%%%%%%%
\section{Behavior Simulation}
\begin{comment}
\begin{itemize}
\item  What functionality does a simulation system -- for our scenario -- need to
    provide in order to allow ECQ?
    \item  How realistic do we need to be ?
 \end{itemize}
\end{comment}
    
As outlined above, our application focus is assistive solutions for \acrshort{pwd}. Therefore, the relevant behaviour simulation requires modelling navigation and assistance of \acrshort{pwd}. Various indoor behaviour modelling tools are already available. For  instance, \cite{huang_indoorstg_2013}, proposed \textit{IndoorSTG} tool to model navigation of customers and shop-assistants in a shopping mall scenario. This was later extended by \cite{li_vita_2016} to the \textit{Vita} tool that allowed individual movement customisation. However, existing navigation behaviour simulations only model healthy individuals and tailoring them to simulate cognitive impairments is either impossible or infeasible. 

Several computational models to realise cognitive deficits are also proposed in literature that are based on different cognitive architectures, such as \cite{manning_magellan_2014, andresen_wayfinding_2016}. However, in these tools disorientation in wayfinding is resulted as random walk which does not reflect real behaviour. Therefore, we have designed our own simulations system \acrshort{simdem}, which aims at generating a more realistic navigation behaviour simulation of \acrshort{pwd} and provides a plausible technical and human assistance simulation. In section \ref{subsection:setup}, we briefly introduce the essential aspects of our simulation model. Readers can refer to \cite{Shaukat2021SimDem} for further details on the \acrshort{simdem}.

In order to allow \acrshort{ecq} evaluation, behaviour simulation must include (i) stakeholder simulation and (ii) assistive technology (AT) simulation. Regarding (i), in our domain, plausible simulation of \acrshort{pwd} and caregivers (i.e., nurses) in indoor environment such as a nursing home is required. We also require to include an AT, such as a smart-watch to evaluate intelligent assistive strategies against the provided ethical value-set.  In the following section, we discuss an example scenario of a nursing home resident diagnosed with dementia to illustrate behaviour simulation and related ethical values in our domain problem. 
\subsection{Sample Scenario}
Mr. Alois is a permanent resident of a nursing home which accommodates people diagnosed with dementia of the Alzheimer type (DAT). Mr. Alois resides in his private room in the nursing home and his daily activities are planned as a daily schedule for instance, going for breakfast at 9:00 am. Today Mr. Alois is scheduled to go for a medical appointment at 01:00 pm. He leaves his room and starts walking to the location of this particular appointment. During his travel, he becomes disoriented and starts moving in the wrong direction. Upon detecting the abnormal behaviour, his smart-watch issues a navigation hint. Assuming this hint to be effective in reorienting Mr. Alois, he returns to the correct path again and eventually reaches desired location without any external/human interference (i.e., staying \textit{autonomous}). After spending required time at medical appointment, Mr. Alois starts travelling back to his personal room. During this travel, he again becomes disoriented. However, this time the smart-watch fails to help him after several attempts. Upon encountering this failure,  smart-watch notifies caregivers so that they can physically help him (i.e., ensuring \textit{beneficence}). This scenario shows, how ethical values (autonomy, beneficence) are connected to system (\acrshort{iat}) actions. Few important aspects of \acrshort{simdem} \cite{Shaukat2021SimDem} to simulate behaviour in the domain of this case study are listed in the following section. 
\subsection{Modeling System Setup}
\label{subsection:setup}

\textit{\acrshort{simdem}} is an agent based simulation model that includes \textit{\acrshort{pwd}}, \textit{nurse} and \textit{smart-watch} agents. Each of these agents have specific roles and abilities to interact with each other within a pre-defined indoor \textit{environment}:

\subsubsection{\textbullet{} PwD Agent:} Represents a nursing home resident. A selected set of \acrshort{pwd} agent parameters are as follows:
\begin{itemize}
    \item \ensuremath{S} is the \textit{schedule} of each patient which contains a set of appointments and each appointment's location and time are also part of the schedule. 
    \item $p_d \in [0,1]$ is the \textit{disorientation level} and models the probability of disorientation arising in above step for a \acrshort{pwd} type agent. 
    \item $p_i \in [0,1]$ refers to the probability that a smart-watch \textit{navigation intervention} will be successful (i.e., helps \acrshort{pwd} to regain orientation) 
\end{itemize}
\subsubsection{\textbullet{} Smart-watch agent:} Each \acrshort{pwd} type agent can be assisted with a smart-watch that can (i) provide navigation hints (ii) remind \acrshort{pwd} of forgotten appointments (iii) call caregivers with \acrshort{pwd} agent's location. Following parameters of the \textit{smart-watch} agent are of special interest here:
\begin{itemize}
    \item \ensuremath{p_{detect} \in [0,1]} represents the \textit{sensor model} of the  smart watch and denotes the probability for each simulation step that disorientation of the monitored PwD is detected by the smart-watch.
    \item \ensuremath{n_{help} \in \mathbb{N}^{*}} denotes the maximum number of consecutive, failed interventions (\ensuremath{n} \ensuremath{\in \mathbb{N}}), before a help intervention is triggered (i.e., a nurse agent is called).
\end{itemize}
\subsubsection{\textbullet{} Nurse agent:} Besides smart-watch assistance, \acrshort{simdem} also models caregiver assistance. A nurse type agent will begin to guide a disoriented \acrshort{pwd} if (i) the nurse agent \textit{perceives} a disoriented patient agent or (ii) the smart-watch agent \textit{calls} the nurse agent to help a patient. To model (i) we used a ray-casting based algorithm where  a nurse agent can only detect a patient within certain radius. In (ii) \acrshort{pwd} agent's exact position is provided to the nurse agent which is then used to track a particular \acrshort{pwd} agent. 

The nurse agent becomes \textit{inactive} when none of the \acrshort{pwd} agents require assistance (i.e., neither a visible disoriented \acrshort{pwd} agent nor a smart-watch intervention to call help). We are aware of the fact that in reality nurses do not simply become inactive and have a lot of tasks to perform. However, this is just a simplification step used in our simulator. 

For now, we assume always \textit{compliant}  patient agents who do not refuse guidance offered by the nurse agent. Non-compliant \acrshort{pwd} agents are planned for the future work to model problematic behaviour such as anger, frustration etc.

\subsubsection{\textbullet{} Environment:} We model the spatial structure of an indoor environment (e.g., a nursing home) as a  two-dimensional orthogonal grid of size \ensuremath{n \times m}. Each grid cell encodes spatial information of the environment such as locations of interest (e.g., dining area, medical clinic, visitation rooms), boundaries (e.g., walls) and other architectural features such as hallway, toilets etc.

\subsubsection{\textbullet{} Spatial Layout:} 
\begin{figure}[!t]
    \centering
    \includegraphics[width=0.6\textwidth]{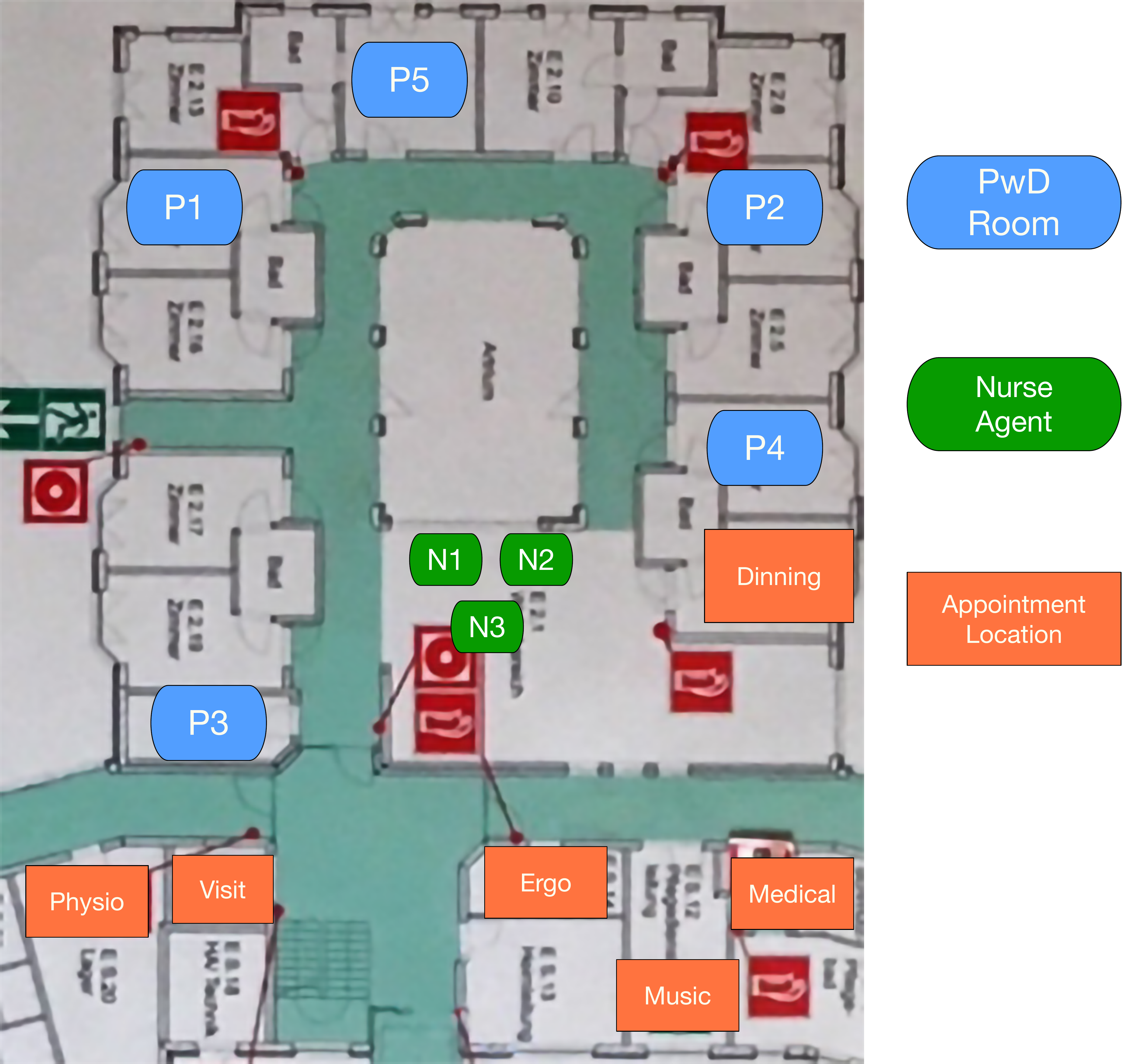}
    \caption{Spatial structure of a nursing home environment showcasing different agents and places of interest}
    \label{fig:outlook}
\end{figure}

Figure \ref{fig:outlook} shows the floor plan of a nursing home simulated using SimDem. This example shows 5 different \acrshort{pwd} agents (\ensuremath{P_1 ,..., P_5}) located in their personal rooms, three nurse type agents (\ensuremath{N_1, N_2, N_3}) located at a common area which is the place where nurse agents reside if they are inactive. Finally, we have some interesting spatial locations  such as dining area, medical clinic, visit room and therapy rooms etc. Note that more \acrshort{pwd} and nurse type agents can be integrated into the environment as per requirements.

%%%%%%%%%%%%%%%%%%%%%%%%%%%%%%%%%%%%%%%%%%%%%%%%%%%%%%%%%%%%%%%%%%%%%%%%%%%%%%%%%%%%%%%%%%%%%%%%%%%%%%%%
\section{Ethical Value Quantification}
From the viewpoint of \acrshort{ecq}, it is now of interest, in how far an assistive strategy (such as the strategy provided by the smart-watch and nursing agent outlined in the previous section), respects set of ethical value preferences. We propose \acrshort{ecq} as an empirical approach to measure the \textit{adherence-level} of an \acrshort{iat} operation against provide ethical value-set. Performing \acrshort{ecq} is a two-step process: (i) defining the \textit{ethical value-set} (ii) defining \textit{value measurement} procedure for each ethical value defined in (i).
\subsubsection{(i) Value Definition:} Depending on the domain, different set of ethical values can be of interest to the stakeholders. For instance, in healthcare, \textit{autonomy} of the patients is of critical importance, whereas in self-driving automobiles context \textit{safety} could be seen as a most prominent ethical value among others. \acrshort{ecq} thus can not predefine a normative set of rules. Rather, the set of relevant ethical values is a design decision.  

An important aspect for defining a value-set is to focus on values that are \textit{observable} by the available sensory information. For instance, \textit{beneficence} towards patients can be observed by measuring the effectiveness of different assistive strategies such as how a smart-watch helps to guide a patient. However, values such as general emotional well-being of the patients might not be observable or derivable with the available information. 
\subsubsection{(ii) Value Measurement:}
Once a value-set is defined, a concise rule for each value is required to decide which actions of the stakeholders or \acrshort{iat} result in violation of that particular value. For example, a physical intervention by the caregivers could be seen as a violation of personal \textit{autonomy} of a patient. 
\subsection{Choice of Values:}
As in this work we are primarily focused on healthcare domain, we chose the most frequently mentioned principles of medical ethics for illustrating use of the \acrshort{ecq} process: (i) Autonomy, (ii) Beneficence, (iii) Justice to provide a proof of concept for our approach. Note that our approach is not limited to these values only and can be applied to stakeholder-defined values.
\subsection{Value Computation:}
The two-step process to obtain \acrshort{ecq} on selected ethical values is as follows: 
\subsubsection{\textbullet{} Value Definition:}
\label{subsub:values}
\begin{itemize}
    \item \textit{Autonomy}: We define autonomy of a patient type agent as the percentage of time a \acrshort{pwd} agent is \textit{not} guided by a nurse agent. 
    \item \textit{Justice}: We confine the broader term of justice to \textit{resource allocation}. We measure how different assistive strategies help caregivers to better allocate their time towards patients. We will refer to this concept as the nurse agent's efficiency.
    \item \textit{Beneficence}: Beneficence of \acrshort{iat} towards patients is a much broader concept. However, in this work, we measure one aspect of it that we call \textit{travel efficiency} that quantifies how efficiently patients reach their appointment destinations as a result of strategy tuning. 
\end{itemize}

\subsubsection{\textbullet{} Value Measurement}
\begin{itemize}
    \item \textit{Autonomy}: Formally, we measure autonomy as: 
    $$\mathit{Autonomy}(\%) =  \left (1 - \frac{ t_{guided}}{t_{total}} \right) \times 100$$ where $t_{guided}$ is the time for which \acrshort{pwd} agent is guided by a nurse agent and $t_{total}$ is the total simulation time.
    \item \textit{Efficiency}: Nurse type agent's efficiency is measured as: 
    $$ \mathit{Efficiency}  (\%) = \left (\frac{t_{inactive}}{t_{total}}\right) \times 100$$ where $t_{inactive}$ is the time for which nurse agent is inactive (i.e., not pursuing or guiding a \acrshort{pwd} agent) and $t_{total}$  is same as above.
    \item \textit{Travel Efficiency (TE)}: Travel efficiency of a \acrshort{pwd} agent is measured as:
    $$\mathit{TE} (\%) = \left (\frac{t_{nominal}}{t_{taken}} \right) \times 100$$ which is the ratio of nominal time required and actual time taken to reach a goal (i.e., appointment location).
\end{itemize}
\subsection{Experiment Design and Results}
\subsubsection{\textbullet{} Experiment Design:}
To perform \acrshort{ecq} for values defined in section \ref{subsub:values}, we setup \acrshort{simdem} environment as follows: 
\paragraph{PwD agents:} For all experiments, we simulated 5 \acrshort{pwd} agents whose home location was as shown in Figure \ref{fig:outlook}. Tuning parameters (section \ref{subsection:setup}) were set as:
\begin{itemize}
    \item \ensuremath{S}: For each simulation run and \acrshort{pwd} agent, 6 unique appointments were selected. Type and location of these appointments were a subset of locations shown in \ref{fig:outlook} (e.g., dining).
    \item \ensuremath{p_d}: five disorientation levels were selected as $0$, $0.25$, $0.50$, $0.75$, and $1$ for all PwD agents.
    \item \ensuremath{p_i}: probability of a successful intervention was set as $0.20$ for each patient.
\end{itemize}
\paragraph {Smart-Watch agents:} Each \acrshort{pwd} agent was equipped with a smart-watch using following parameters:
\begin{itemize}
    \item \ensuremath{p_{detect}}: was set as $0.5$ and $0.2$ for all experiments.
    \item \nhelp{}: six values of \nhelp{} were experimented from $0$ to $5$.
\end{itemize}
\paragraph{Measurement:} We analyse \acrshort{ecq} against different assistive strategies (i.e., \nhelp{}). As we increase the \nhelp{} value, calling the caregiver on a disoriented patient agent is delayed. For instance, if \nhelp{} $=1$, nurse agent will be called after one failed \textit{navigation intervention} by the smart-watch. Similarly, \nhelp{} $=5$ means that five navigation attempts will be made by the smart-watch before informing a nurse agent. 
\begin{figure}[t]
    \centering
    \includegraphics[width=0.9\textwidth]{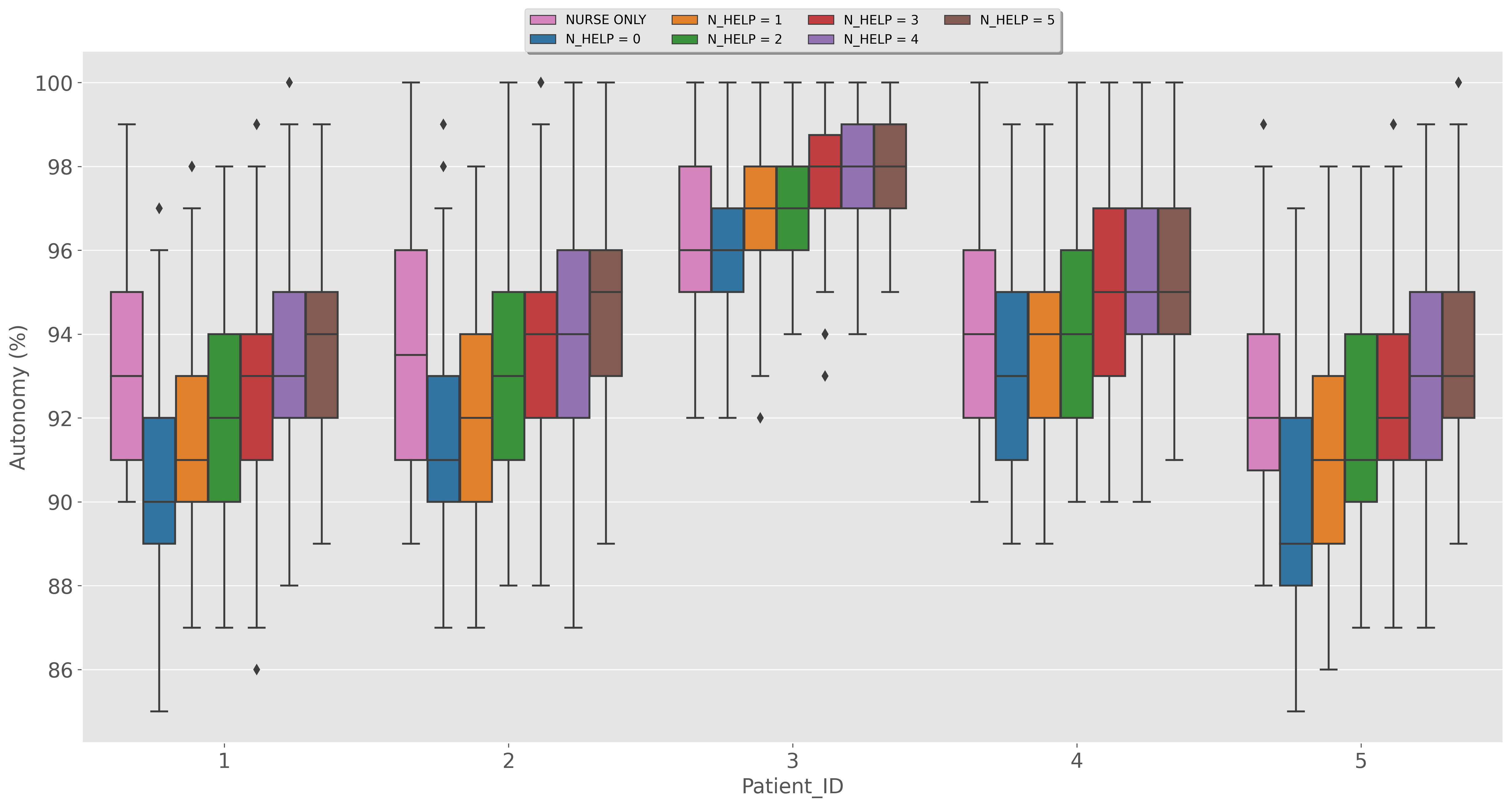}
    \caption{Patient's autonomy versus (\nhelp{})}
    \label{fig:autonomy}
\end{figure}
\subsubsection{\textbullet{} Results:} Figure \ref{fig:autonomy} shows the effect of tuning \nhelp{} on the autonomy of each patient. As we increase \nhelp{}, guidance to the \acrshort{pwd} is provided less often and that can be said to increase patient agent's autonomy. This effect is visible for all of the patient agents. We can also observe that without any smart-watch assistance, \acrshort{pwd} agents are less autonomous as compared to larger \nhelp{} values (i.e., \nhelp{} $=4,5$). 

\begin{figure}[t]
    \centering
    \includegraphics[width=0.9\textwidth]{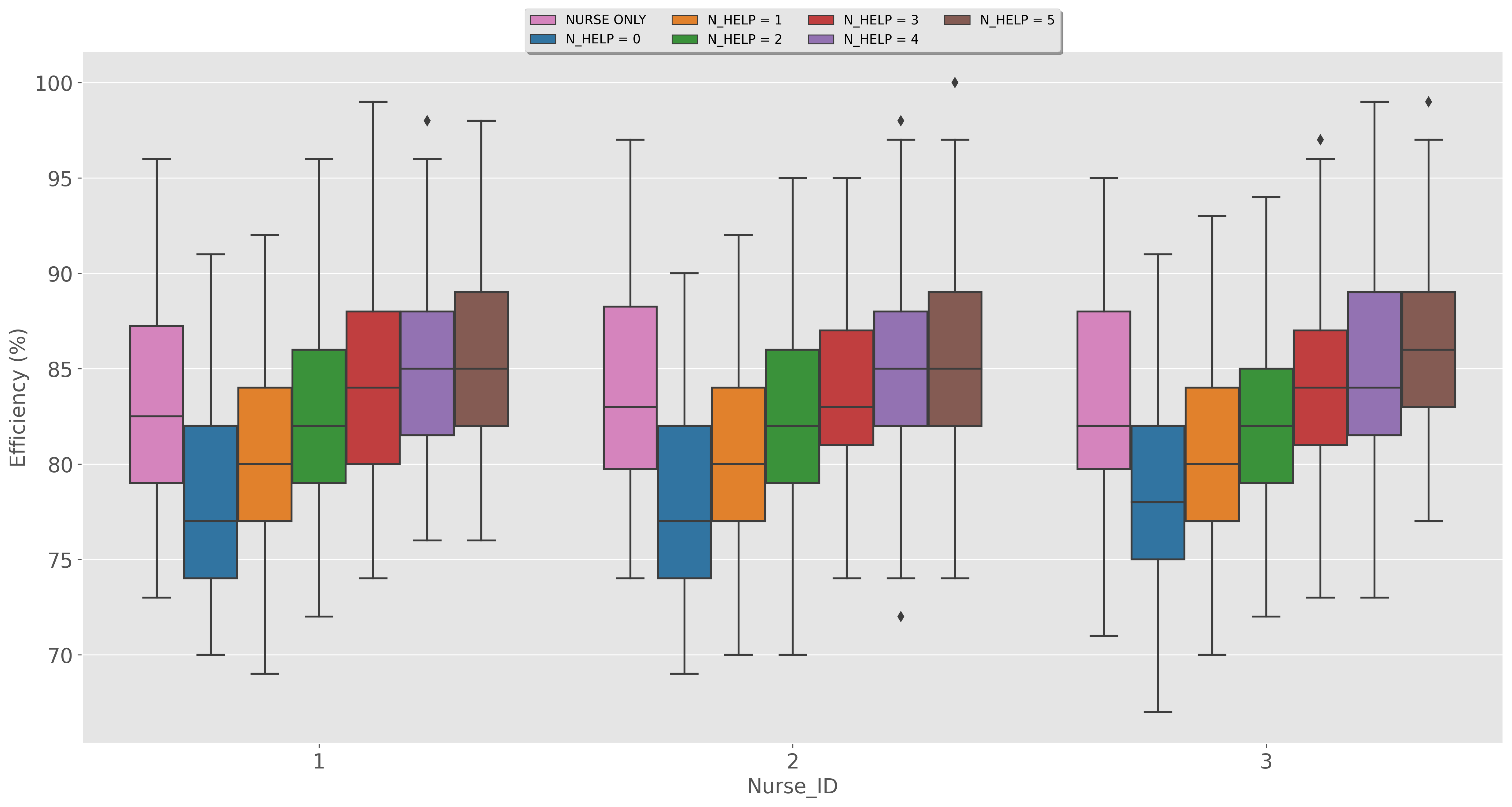}
    \caption{Nurse's efficiency versus different values of assistive strategies (\nhelp{})}
    \label{fig:justice}
\end{figure}

Figure \ref{fig:justice} depicts how efficient (i.e., stays inactive) the nurse agents become against \nhelp{}. We can clearly observe that all of the nurse agents have more available time if we increase the \nhelp{} value. This is due to the fact that the nurse agents are called more often for lower values of \nhelp{}. For instance, at \nhelp{} $=0$ a nurse is called as soon as a patient agent gets disoriented.

\begin{figure}[!b]
    \centering
    \includegraphics[width=0.9\textwidth]{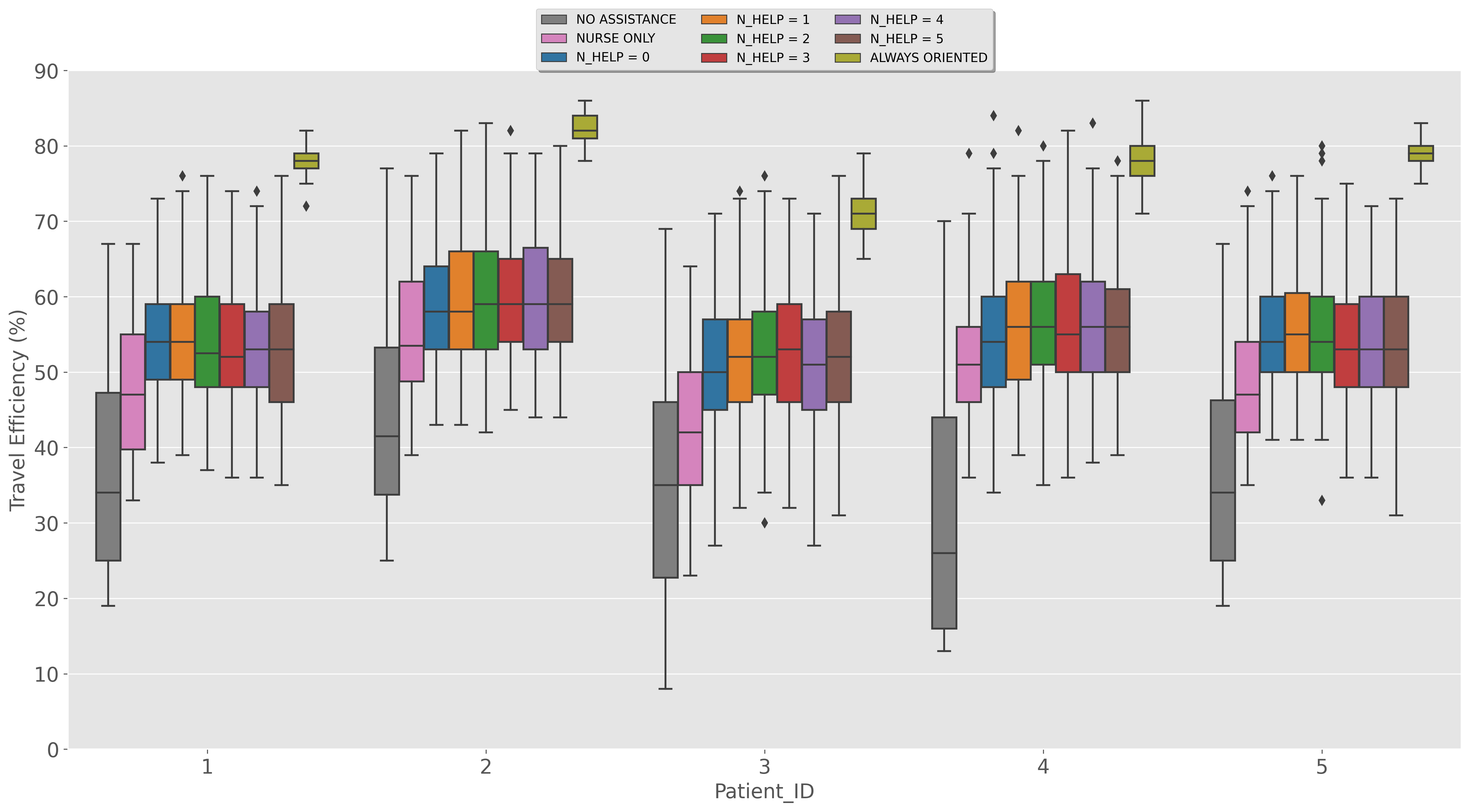}
    \caption{Patient's travel efficiency versus different values of assistive strategies (\nhelp{})}
    \label{fig:beneficience}
\end{figure}

Moreover, without any smart-watch, nurse agents stay inactive most of the time as they are generally unaware of any disoriented patient agent unless they perceive them within their visible area. We might conclude that having a smart-watch for the \acrshort{pwd} agents does not bring any benefit in terms of nurse type agent's efficiency.  However, from the perspective of \acrshort{pwd} agents, having a smart-watch assistance definitely provides \textit{beneficence} towards patient agents. This is shown in the Figure \ref{fig:beneficience} where we measure travel efficiency of \acrshort{pwd} agents against \nhelp{}. First thing to notice is that the \te{} increases significantly as we provide any kind of assistance with or without smart-watch. Secondly, providing smart-watch assistance to the \acrshort{pwd} agents, in general, increases their \te{}, however, changing \nhelp{} does not have a significant effect on the \te{}. Note that, \te{} does not reach $100 \%$ even when \acrshort{pwd} agent is always oriented. This behaviour is a feature of our model \acrshort{simdem} to model postural stability and localization error present in the real-world. 

Comparing autonomy (Figure \ref{fig:autonomy}) and \te{} (Figure \ref{fig:beneficience}) shows that \acrshort{pwd} agents can still travel efficiently if we delay caregiver's help (i.e., increasing \nhelp{}) and at the same time increase their autonomy by choosing larger \nhelp{} values. Moreover, comparing Figure \ref{fig:beneficience} and \ref{fig:justice} shows that increasing the \nhelp{} value improves nurse agent's efficiency while not having a significant effect on \acrshort{pwd} agent's \te{}.

\subsubsection{\textbullet{} Discussion:}
The results obtained from this simple model illustrate, how a simulation-based approach to \acrshort{ecq} is able to quantify the \textit{adherence} of specific assistive strategy to a set of ethical values.  We can clearly see the effect of different values of \nhelp{} on the value quantification, and we also see the value \textit{trade-off} between \textit{autonomy}, \textit{justice} and \textit{beneficence}.  Therefore, this simulation-based method to \acrshort{ecq} provides a flexible mechanism to understand the interaction between assistive strategies and ethical values.
%%%%%%%%%%%%%%%%%%%%%%%%%%%%%%%%%%%%%%%%%%%%%%%%%%%%%%%%%%%%%%%%%%%%%%%%%%%%%%%%%%%%%%%%%%%%%%%%%%%%%%%%
\section{Conclusion and Future Work}
\begin{comment}
- Defined ethical values are an initial effort towards solving the problem, however, approach can be adapted to measure other ethical values in different domains \\
- We can not be sure if the result would actual stay true in real scenarios, therefore further investigation is required , however, our approach still proceeds in the right direction
\end{comment}
We presented a novel approach, \textit{ethical compliance quantification}, to quantify adherence-level of assistive systems for given ethical value-set. We argued that regardless of integration mechanism, we still need to \textit{analyse} how well \acrshort{iat} performs to implement ethical rules. As a proof of concept, we utilised biomedical ethical \textit{value set}, however, we believe our approach can be adapted to stakeholder defined rules. 

Of course, our definitions of ethical values (e.g., autonomy) are \enquote{over simplified}. For instance, the concept of autonomy might not be reduced to the simple rule of caregiver guidance. In this regard, we argue that purpose of this work is not to perfectly define ethical values but to present an approach that can use defined values and analyse how \acrshort{iat} \textit{comply} against these values. As future work, we therefore plan to add and refine such value-sets.

The simulation based \acrshort{ecq} provides a mechanism for empirical research towards \acrshort{ecq}. Note that simulation-based \acrshort{ecq} also allows to analyse situations that would be \textit{unethical} in the real world, such as providing no help at all and compromising on \textit{safety} of PwD (so that we are able to understand the real effect of assistance). 
Also, few simplifications in our behaviour simulator \acrshort{simdem} are also present. Such as \acrshort{pwd} agents that always comply with the nurse agents. Our planed work, therefore, also include integrating problematic behaviour of the \acrshort{pwd} agent such as aggression, frustration and refusing provided help by the nurse agents. However, our main goal here is to not report a perfect simulation tool as developing such tools is a continuous process. Rather, we focus on how to use such a behaviour simulation for \acrshort{ecq}.

\section{Acknowledgment}
This research is funded by the German Federal Ministry
of Education and Research (BMBF) as part of the EIDEC
project (01GP1901C), and by the European Union (EU,
EFRE) as part of the SAMi project (TBI-1-103-VBW-035).

%%%%%%%%%%%%%%%%%%%%%%%%%%%%%%%%%%%%%%%%%%%%%%%%%%%%%%%%%%%%%%%%%%%%%%%%%%%%%%%%%%%%%%%%%%%%%%%%%%%%%%%%

\bibliographystyle{ieeetr}
\bibliography{bib}
\end{document}